\documentclass[%
 reprint,
superscriptaddress,
 amsmath,amssymb,
 aps,
pra,
floatfix,
]{revtex4-2}

\usepackage{orcidlink}
\usepackage{graphicx}
\usepackage{dcolumn}
\usepackage{bm}
\usepackage{siunitx}
\usepackage[dvipsnames]{xcolor}
\usepackage{ulem} 
\usepackage{lipsum}
\usepackage{textcomp}
\usepackage{upgreek}

\DeclareSIUnit\gauss{G}

\begin{document}

\preprint{APS/123-QED}

\title{Narrow-line magneto-optical trap of titanium atoms}

\author{Rowan Duim~\orcidlink{0000-0003-1970-3892}}
\affiliation{Department of Physics, University of California, Berkeley, CA 94720}
\affiliation{Challenge Institute for Quantum Computation, University of California, Berkeley, CA 94720}

\author{Scott Eustice~\orcidlink{0000-0003-1102-2400}}
\altaffiliation[Present address: ]{Joint Quantum Institute, National Institutes for Standards and Technology, College Park, MD 20742}
\affiliation{Department of Physics, University of California, Berkeley, CA 94720}
\affiliation{Challenge Institute for Quantum Computation, University of California, Berkeley, CA 94720}

\author{Jackson Schrott}
\affiliation{Department of Physics, University of California, Berkeley, CA 94720}
\affiliation{Challenge Institute for Quantum Computation, University of California, Berkeley, CA 94720}

\author{Hiromitsu Sawaoka~\orcidlink{0000-0003-4657-1303}}
\affiliation{Department of Physics, University of California, Berkeley, CA 94720}
\affiliation{Challenge Institute for Quantum Computation, University of California, Berkeley, CA 94720}

\author{Dan M.\ Stamper-Kurn~\orcidlink{0000-0002-4845-5835}}
\affiliation{Department of Physics, University of California, Berkeley, CA 94720}
\affiliation{Challenge Institute for Quantum Computation, University of California, Berkeley, CA 94720}
\affiliation{Materials Sciences Division, Lawrence Berkeley National Laboratory, Berkeley, California 94720}
    
\date{\today}

\begin{abstract}
We realize narrow-linewidth magneto-optical traps of $^{46}$Ti, $^{48}$Ti and $^{50}$Ti atoms based on a \qty{1040}{\nm}-wavelength transition, cooling atoms to a minimum temperature in one dimension of $T_z=\qty{990(20)}{\nano \kelvin}$ and a three-dimensional temperature of $T_\mathrm{3D}=\qty{1.28(7)}{\upmu \kelvin}$.
Atoms are pre-cooled in a broad-line magneto-optical trap and then transferred with about 25\% efficiency to the narrow-line trap. We operate the narrow-line trap in two stages over 85 ms.
First, a single cooling beam, blue-detuned from the narrow-linewidth atomic resonance, optically pumps and traps the atoms on a two-dimensional surface where the Zeeman shift from the applied spherical quadrupole magnetic field brings the light nearly to resonance. Second, four additional beams, counter-propagating in the transverse directions, cool and compress the atoms in all dimensions.
The high magnetic moment of the laser cooling state makes the dynamics of the narrow-line titanium trap similar to those of other magnetic atoms.
We measure the lifetime of the excited state of the transition to be $\tau=\qty{8.2(9)}{\upmu \second}$, indicating a transition linewidth of  $\gamma/2\pi =20(2)\ \unit{\kHz}$, and also measure isotope shifts on the narrow-line transition.
We use Stern-Gerlach separation on the ultracold Ti gas to measure the $m_J$-distribution in the narrow-line magneto-optical trap, finding over 98\% of the atoms in the stretched spin state.
\end{abstract}

\maketitle

Narrow-linewidth transitions have proven a powerful tool for the laser cooling of atoms, ions and molecules.
The velocity selectivity of a narrow line gives rise to low Doppler temperatures and facilitates high phase-space density magneto-optical traps (MOTs) \cite{katori_magneto1999, vogel_narrow1999, kuwamoto_intercombination1999, frisch_narrow_2012, mehling_narrowline2025}.
For atoms subject to a confinement potential, a linewidth that is small compared to the trap frequency allows for resolution of motional states and sideband cooling to the ground state \cite{wineland_laser1987, diedrich_laser1989, monroe_resolved1995}.
Laser cooling of non-alkali species, such as alkaline-earths and lanthanides, has been driven by their rich electronic structures offering properties such as long-lived metastable states, ultra-narrow clock transitions, and large internal angular momenta and magnetic moments \cite{honda_magneto-optical1999, lu_trapping2009}; the availability of narrow laser cooling lines has accelerated these efforts and helped to establish these atoms as leading platforms in atomic physics.

\begin{figure*}
    \centering
    \includegraphics[scale=0.85]{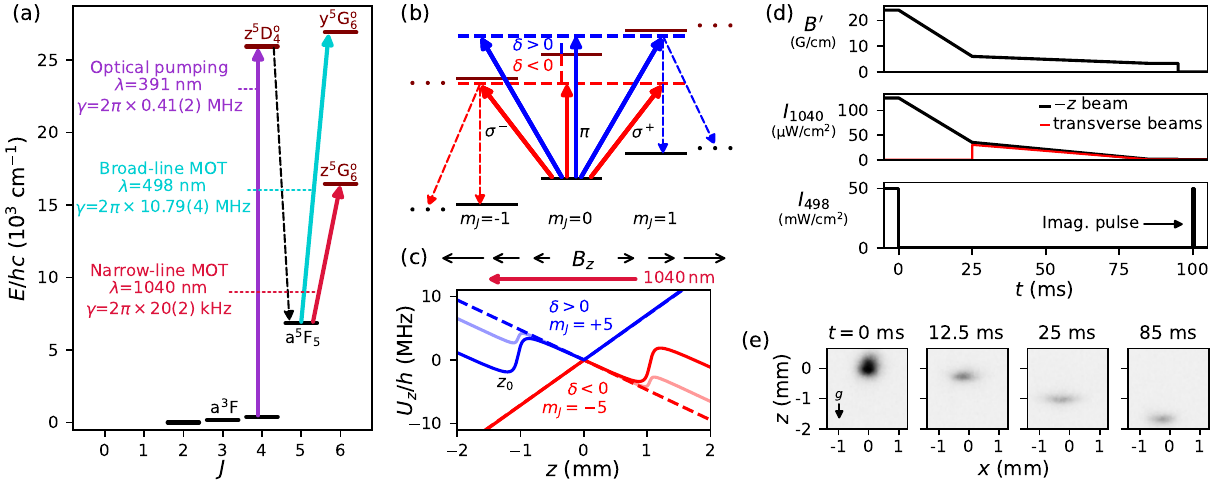}
    \caption{(a) A reduced energy level diagram of Ti is shown, with even- (odd-) parity levels relevant to laser cooling shown as black (maroon) lines. 
    The optical pumping, broad linewidth laser cooling, and narrow linewidth laser cooling transitions are indicated by purple, cyan and crimson arrows, while the $\mathrm{y^5D_4^o\to a^5F_5}$ decay is shown with a dashed black arrow.
    (b) A simplified diagram of the interaction of the narrow-line laser with different $m_J$ sublevels.
    In the presence of a finite magnetic field and with the laser frequency blue-detuned from resonance, $\sigma^+$ ($\sigma^-$) transitions are brought closer (farther) from resonance, leading to optical pumping towards the $m_J=+J$ sublevel.
    When the laser frequency is red-detuned, the role of each polarization flips and the atom is pumped to the $m_J=-J$ sublevel.
    (c) The potential energies along the $z$ axis for stretched-state atoms in a single blue- or red-detuned $- \hat{z}$ beam narrow-line MOT are shown in their respective colors.
    Dashed lines show the magnetic potential alone, while light (dark) solid lines show the combined potential generated by both magnetic forces and optical forces at a saturation parameter of $s_z$=1 ($s_z$=5).
    The linear magnetic field gradient along the $z$ axis is indicated with black arrows above, and the direction of the left-hand circularly polarized $\lambda=1040$\,nm beam, indicated in crimson, is the same for the red- and blue-detuned cases.
    (d) Experimental sequence for loading atoms into the narrow-line MOT, showing the values of the magnetic field gradient $B'$, intensity of the \qty{1040}{\nm} vertical and transverse beams $I_{1040}$, and intensity of the \qty{498}{\nm} broad-line MOT beams $I_{498}$ versus time.
    The detuning of the \qty{1040}{\nm} beams is fixed throughout the sequence, with $\delta\approx23\,\gamma$ as the typical operating condition.
     (e) Fluorescence images of atoms being loaded into the narrow-line MOT from the magnetic trap are shown, with 10 individual realizations of the MOT averaged for each image.
     The direction of gravity is shown in the first image.
     }
    \label{fig:E-levels-nMOT-cartoon}
\end{figure*}

Neutral titanium further enriches the atomic physics toolbox, notably offering both nonzero electronic orbital angular momentum and small magnetic moment in the ground state, a telecommunications-band clock transition, and five stable isotopes~\cite{eustice_optical_2023}.
Ti has no cycling transition out of the ground $\mathrm{a^3F}$ term, but can be optically pumped into the metastable $\mathrm{3d^3(^4F)4s\,a^5F_5}$ state, which possesses two nearly cycling transitions suitable for laser cooling~\cite{eustice_optical_2023}: the broad $\mathrm{a^5F_5 \rightarrow y ^5G_6}$ transition, at the wavelength of 498 nm and a linewidth of $\gamma/{2\pi}=\qty{10.79(4)}{\MHz}$, and the narrow $\mathrm{a ^5F_5 \rightarrow z^5G_6^o}$ transition, at the wavelength of 1040 nm and a linewidth of $\gamma/{2\pi} = \qty{20(2)}{\kHz}$ (as measured in this work).
See Fig~\ref{fig:E-levels-nMOT-cartoon}(a) for a simplified energy level diagram.
MOTs operating on the broad line have been reported for all five stable isotopes of Ti~\cite{eustice_magneto-optical_2025,schrott_hyperfine2026}.

Here, we report second-stage magneto-optical trapping of Ti on the narrow-line transition.
The  $\mathrm{a^5F_5}$ laser-cooling state of Ti has a large magnetic moment $\mu\simeq 7\mu_\mathrm{B}$, comparable to the ground states of Eu (7\,$\mu_\mathrm{B}$), Dy (10\,$\mu_\mathrm{B}$) and Er (7\,$\mu_\mathrm{B}$).
This large magnetic moment modifies the behavior of narrow-line MOTs as compared to broad-line MOTs~\cite{berglund_narrow_2008, lu_strongly_2011, frisch_narrow_2012, maier_narrow_2014,miyazawa_narrow2021}.  
Following the example demonstrated for narrow-line cooling of other strongly magnetic atoms, starting from a broad-line MOT, we implement a second cooling stage wherein the optical scattering force of a single, blue-detuned $\lambda=\qty{1040}{\nm}$ beam balances the force of the magnetic trap, forming a local potential minimum where Doppler cooling on the narrow line occurs.
Atoms from the broad-line MOT are optically pumped into weak-magnetic-field-seeking states, held initially within the magnetic trap formed by the applied spherical quadrupole magnetic field, and then are transferred over 10s of ms into the single-beam narrow-line MOT.
After ramping down the optical power and magnetic-field gradient, which reduces the atom temperature along the axis of the single beam, transverse beams are added to cool and compress the gas in three dimensions.
We achieve a narrow-line MOT for $^{48}$Ti with a three-dimensional temperature of $T_{\mathrm{3D}}=\qty{1.28(7)}{\upmu \kelvin}$, and a one-dimensional temperature as low as $T_{\mathrm{z}}=\qty{990(20)}{\nano \kelvin}$.

The availability of trapped Ti atoms at $\upmu$K-scale temperatures allows us to characterize the narrow-line transition.
By shelving atoms in the excited $\mathrm{z^5G_6^o}$ state and probing their return to the $\mathrm{a^5F_5}$ state by fluorescence on the broad-line laser cooling transition, we measure the $\mathrm{z^5G_6^o}$ state lifetime to be $\tau=\qty{8.2(9)}{\upmu \s}$.
We apply a Stern-Gerlach analysis to measure the magnetic sublevel distribution of optically pumped and depumped gases of Ti, and use such analysis to pinpoint the zero-magnetic-field atomic resonance frequency on the narrow-line transition.
Further, by realizing narrow-line MOTs also of the $^{46}$Ti and $^{50}$Ti isotopes, we measure transition isotopes shifts.  

The paper is organized as follows: Sec.~\ref{sec:theory} describes the principle of the magnetic atom narrow-line MOT and the 1D analytical theory of the system.
In Sec.~\ref{sec:optimization} we detail the experimental procedure and optimization of the narrow-line MOT.
Sec.~\ref{sec:spectroscopy} describes the Stern-Gerlach analysis of the ultracold Ti gas, the determination of isotope shifts, and the measurement of the excited state lifetime. 

\section{Operating principle of the Ti narrow line MOT}
\label{sec:theory}
Magneto-optical traps based on narrow atomic transitions typically suffer from small capture velocities and volumes owing to the confluence of weak optical scattering forces and narrow frequency ranges over which atoms are optically active. 
Consider atoms moving in a spherical quadrupole field $\mathbf{B}=B'(z\mathbf{\hat{z}}-\rho\bm{\hat{\rho}}/2)$ and interacting with laser light.
Here, $\mathbf{B}$ is expressed in cylindrical coordinates with $\bf{\hat{z}}$ denoting the spherical quadrupole field axis and $\bm{\hat{\rho}}$ the transverse radial direction.
We assume the magnetic field gradient $B^\prime>0$.
A MOT is formed utilizing a $J_g \rightarrow J_e = J_g+1$ transition with optical wavevector $k = 2 \pi/\lambda$, natural linewidth $\gamma$, and a magnetic moment difference $\Delta\mu = (g_e J_e - g_g J_g) \mu_B$ between the stretched $m_J$ states of the excited and ground states; here, $g_e=4/3$ and $g_g=7/5$ are the Land\'{e} $g$-factors of the excited and ground state of the laser cooling transition, respectively, with specific values given for both narrow- and broad-line MOTs of Ti, and $\mu_B$ is the Bohr magneton.
We consider that the MOT light is detuned by $\delta$ from the zero-magnetic-field resonance frequency. 

Given this information, we can characterize the MOT with two length scales.
We define an active length $z_B = \hbar \gamma / \Delta\mu B^\prime$ as the length along the axial direction of the trap over which the unsaturated transition is Zeeman shifted across the transition linewidth.
Multiplying by the maximum radiation-pressure force, $F_\gamma = \hbar k \gamma/2$, we obtain $U = F_\gamma z_B$, an estimate of the trap depth, and, with $U = m v_c^2/2$, an estimate of the trap capture velocity $v_c$ where $m$ is the atomic mass.
A second characteristic length $z_\delta =  z_B \, |\delta|/\gamma$ is the axial distance from the trap center where the laser-cooling light is shifted onto resonance on the stretched-state transition.
This length defines a capture length (or area, or volume) of the MOT.
We define these length scales considering just motion along the MOT axis.
Similar length scales pertain to displacement in the transverse direction and define the three-dimensional dynamics of the MOT.

Let us compare length scales of the broad- and narrow-line MOTs for $^{48}$Ti.
The broad-line MOT, with a typical gradient of $B^\prime = 10$ G/cm, has an active length of $z_B = 0.8$ cm.
A large capture velocity $v_c = 93$ m/s allows the broad MOT to be loaded directly from a Ti-sublimation source.
The radiation pressure force is far larger than the magnetic force, with $F_\gamma/F_B = 7 \times 10^3$.
Thus, with the MOT operating with a typical red detuning ($\delta<0$) of several linewidths, magnetic forces are negligible throughout the MOT volume.

In contrast, the narrow-line MOT at the same magnetic gradient has an active length of $z_B = \qty{14}{\upmu m}$ and a capture velocity of $v_c = \qty{0.1}{m/s}$.
Owing to the small capture velocity, we choose to load a narrow-line MOT with a Ti gas that is initially laser cooled and trapped by a broad-line MOT, rather than loading it directly from the Ti-sublimation source.
To encompass the entire volume of the broad-line MOT, which has a radius on the order of \qty{100}{\upmu m}, we operate the narrow-line MOT at a detunings of several tens of linewidths.
Additionally, at the settings considered here, the radiation pressure force on the narrow-line transition is comparable to the magnetic force, with $F_\gamma/F_B = 6$.
Thus, in a narrow-line MOT the magnetic force dominates the motion of ground-state atoms outside of a shell, with a thickness of just tens of micrometers, along the surface $S_\delta$ (which intersects the $z$ axis at $\pm z_\delta$) where the laser-cooling light is Zeeman shifted into resonance.

While several methods exist to increase the capture volume and velocity of narrow-line MOTs~\cite{vogel_narrow1999, chaneliere_three2008}, our approach is to operate the narrow-line MOT at a large \textit{blue} detuning from resonance.
As illustrated in Fig.~\ref{fig:E-levels-nMOT-cartoon}(b), in regions with a non-zero magnetic field, Zeeman shifts tend to bring $\sigma^+$ ($\sigma^-$) optical transitions closer to resonance for blue (red) detuned light (so long as the Zeeman shifts do not far exceed the detuning magnitude).
Thus, the effect of blue-detuned light is to pump atoms optically into the weak-field-seeking stretched state ($m_J = +J_g$) while that of red-detuned light is to pump atoms optically into the strong-field-seeking stretched state ($m_J = -J_g$).  
This optical pumping  relies on few photon scattering events, and thus can occur even when atoms are near the quadrupole field zero and not strongly interacting with the laser-cooling light.
For the narrow-line MOT, in which magnetic forces dominate over radiation pressure  over much of the experimental volume, one finds that blue-detuned MOT light provides the advantage that the optically pumped atoms are now magnetically trapped within the MOT's spherical quadrupole field.
As described below, the blue-detuned narrow-line MOT, located on a narrow shell about $S_\delta$, is loaded by gradually collecting and cooling these magnetically trapped atoms.
In contrast, in the case of red detuning, atoms are optically pumped into magnetically anti-trapped states.
Atoms initially at the magnetic trap center are accelerated to high velocities, exceeding $v_c$ by the time they reach $S_\delta$, and cannot be captured by the radiation pressure force.

For $|\delta|\gg\gamma$ we can then reduce the relevant atomic structure to a two-level system in which only the $m_J=5\to6$ transition is driven.
Outside of the surface $S_\delta$, the laser field becomes locally red detuned from this transition, enabling Doppler cooling.
In order to create a point of stable equilibrium outside of $S_\delta$, realizing trapping in addition to local Doppler cooling, an outward optical scattering force must oppose the inward restoring force of the magnetic gradient.
For a blue-detuned beam propagating along the $z$ axis, a force away from the quadrupole field zero is achieved with left-hand circular polarized light.

Fig.~\ref{fig:E-levels-nMOT-cartoon}(c) shows the effective potential along  the $z$ axis determined by integrating the position-dependent force experienced by an optically pumped atom from the combination of magnetic forces, radiation pressure, and gravity.
Local potential minima, near the position where the downward ($- \bf{\hat{z}}$) propagating laser-cooling light is Zeeman shifted into resonance, are seen both for the cases of red- and blue-detuned light.
The figure also illustrates the effect of saturation: in the case of $s_z = I_z/I_\mathrm{sat} > 1$, with $I_z$ being the intensity of the $z$ axis beam and $I_\mathrm{sat} = \qty{2.3}{\upmu W/\cm^2}$ the saturation intensity on the stretched-state transition, one finds the active length of the narrow-line MOT $z_B$ is increased (by the factor $\sqrt{s_z+1}$) and also, correspondingly, the potential depth at the local minimum is increased.
In our work, we strongly saturate the narrow-line transition at early times to facilitate the initial loading of atoms into the narrow-line MOT, and then reduce the beam intensities to $s_z\simeq 1$ to achieve the lowest MOT temperatures. 
All saturation parameters in this work are defined for single beams on the stretched transition and have a fractional uncertainty of 11\%.

\section{Operation and Optimization of the Narrow-line MOT}
\label{sec:optimization}

The Ti MOT apparatus has been described in previous work~\cite{eustice_magneto-optical_2025}. 
We use an Yb-doped fiber laser to generate \qty{0.5}{\watt} of light at the wavelength of $\lambda=\qty{1040}{\nm}$. 
The laser frequency is locked in a tunable manner to an ultra-low expansion (ULE) glass optical cavity via the offset-sideband Pound-Drever-Hall (PDH) technique. 
From residual frequency noise, we judge the laser linewidth to be below 8.5\,kHz; however, such measurements are in-loop and not entirely reliable. 
Alternatively, we can assess the linewidth from the temperature of the laser-cooled atoms: From the proximity of the lowest observed axial temperature of $\qty{990(20)}{\nano \kelvin}$ to the Doppler temperature limit defined as ${\hbar\gamma}/{2k_\mathrm{B}}=\qty{470}{\nano \kelvin}$ as well as the recoil limit $(\hbar k)^2/2m=\qty{180}{\nano \kelvin}$, we infer that the laser linewidth is not appreciably larger than the transition linewidth.  
Narrow-line cooling beams with $1/e^2$-intensity beam diameters of 10 mm (5 mm) in the vertical (horizontal) directions are combined with the broad-line laser-cooling beams via dichroic beamsplitters.  
Dual-wavelength $\lambda/4$ waveplates are used to set the polarization of all narrow-line cooling beams to be left-hand circular.

Our procedure for cooling Ti in a narrow-line MOT is described in Fig.~\ref{fig:E-levels-nMOT-cartoon}(d).
Briefly, a broad-line MOT of Ti atoms is loaded initially from an optically pumped Ti-sublimation source as described in Ref.~\cite{eustice_magneto-optical_2025}.
A single vertical blue-detuned narrow-line MOT beam is sent into the MOT from above (along $-\bf{\hat{z}}$) during the entire 500-ms-long initial loading stage.
We then switch off the broad-line cooling beams (at time $t=0$ in Fig.~\ref{fig:E-levels-nMOT-cartoon}(d)), maintaining the same $B^\prime =$ \qty{24}{G/cm} field gradient as used in the initial MOT.
The 391-nm-wavelength light used to optically pump the atomic beam is also extinguished at this time.
Atoms are pumped by narrow-line light into into the magnetically trapped stretched-$m_J$ state and, over the next $\tau_\mathrm{1D} = 25$ ms, collected within a one-beam narrow-line MOT.
The vertical-beam light intensity $I_z$ and field gradient $B^\prime$ are ramped to lower values during this stage.
Next, we switch on the four horizontal narrow-line MOT beams, with identical detuning as the vertical beam and intensities $I_x$, and cool atoms for a time $\tau_\mathrm{3D}=\qty{60}{\ms}$ in a three-dimensional, 5-beam narrow-line MOT, while further ramping down the light intensity and field gradient.
Finally, we measure the atom number and temperature by switching off all light beams and the field gradient, allowing the atoms to expand freely for a variable time of flight, and then measuring their distribution by fluorescence imaging on the \qty{498}{\nm} wavelength transition.  

\subsection{Optimization of the single-beam MOT}
\label{sec:vertical}

\begin{figure}[t]
    \centering
    \includegraphics{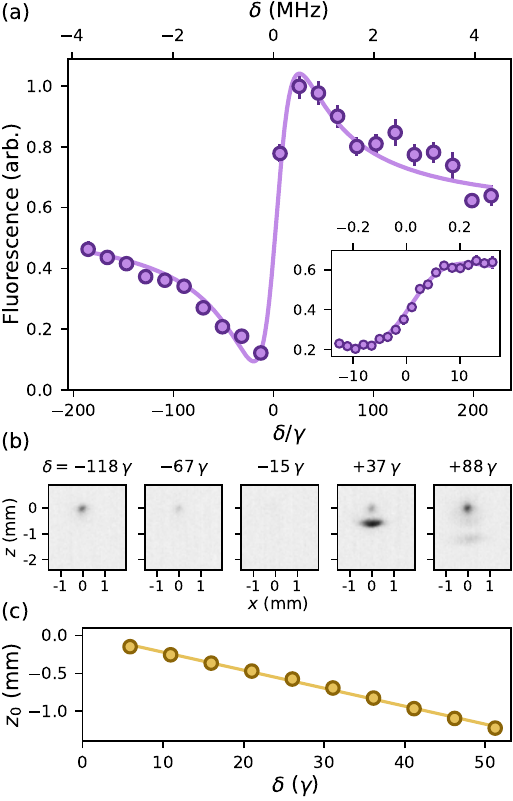}
    \caption{(a) Fluorescence on the $\lambda=498$\,nm line, proportional to the number of atoms in the magnetic trap, is plotted as a function of the applied $\lambda=\qty{1040}{\nm}$ laser detuning indicated in linewidths (bottom axis) and MHz (top axis).
    The inset shows a narrower range of detunings.
    A dispersive lineshape  (light purple line; see Eq.\ \ref{eq:fit}) fits the data.
    Detuning is calibrated to the resonant frequency as measured by Stern-Gerlach spectroscopy.
    Error bars show one standard deviation in shot-to-shot noise; error bars not visible are smaller than the marker size.
    (b) Fluorescence images at several detunings of the narrow line beam, for final saturation parameter of the vertical narrow-line beam of $s_z\approx 90$ and magnetic field gradient $B'=\qty{6}{\gauss / \cm}$.
    The images show depletion of the magnetic trap at red detuning, and population enhancement with narrow-line MOT formation at blue detuning.
    (c) Position of the single-beam narrow-line MOT center below the quadrupole field zero versus detuning, with a fit line.
    The final magnetic field gradient is $B'=\qty{6}{\gauss / \cm}$ and the final saturation parameter of the narrow line beam is $s_z\approx80$.
    } 
    \label{fig:optical_pumping}
\end{figure}

We begin by examining the process of loading the single-beam narrow-line MOT starting from a broad-line MOT of Ti.
We focus on MOTs of the most abundant isotope, $^{48}$Ti, unless otherwise noted.
We first consider the process of optically pumping atoms into magnetically trappable states.
For this, after switching off the broad-line MOT, we expose atoms to a single vertical beam of narrow-line cooling light, at a fixed intensity of $I_z = \qty{135}{\upmu \watt/\centi\meter^2}$ ($s_z\approx 60$), and a variable optical frequency that is held constant over 25 ms of exposure.
We then count the total remaining atom number through fluorescence imaging, with results shown in Fig.~\ref{fig:optical_pumping}(a).
For red or blue detunings beyond several hundred linewidths from the zero-field resonance, the gas is unaffected by the optical pumping light and we observe a frequency-independent number of atoms trapped in the spherical quadrupole magnetic trap.
At closer detuning, we observe a depletion of trapped atoms for the case of red detuning and an increase of trapped atoms for the case of blue detuning.
These observations confirm the tendency of red-detuned light to pump atoms into strong-field-seeking states that are expelled from the magnetic trap, and the tendency of blue-detuned light to pump atoms into weak-field-seeking states that are retained in the trap.
We note that we observe similar atom numbers independent of the polarization of the single incident beam (linear or circular of either handedness), confirming that the pumping into weak- or strong-field-seeking states is caused by the energetics of Zeeman shifts rather than by polarization selection rules.

The observed variation in trapped atom number $N$ with optical frequency $\omega$, where $\omega$ is measured with respect to the stable resonance frequency of our ULE cavity, provides a convenient, albeit somewhat inaccurate, means of measuring the narrow-line resonance frequency.
We fit the data such as those in Fig.~\ref{fig:optical_pumping}(a) to an empirical dispersive lineshape of the form
\begin{equation}
    N(x) = A \frac{x}{1 + x^2} + B
    \label{eq:fit}
\end{equation}
where $x = (\omega - \omega_0)/\Gamma$. Fit constants $\omega_0$, $\Gamma$, $A$ and $B$ correspond to transition center frequency, effective linewidth, amplitude of atom number enhancement, and background atom number. We find a resonance frequency $\omega_0$ that is systematically offset by about 3$\gamma$ from the more accurate value found through a form of Stern-Gerlach spectroscopy (see Appendix~\ref{app:SG}). In Fig.~\ref{fig:optical_pumping}, the frequency axes are expressed in terms of the detuning $\delta$ and centered at $\delta=0$ as determined by Stern-Gerlach spectroscopy. 
As discussed below, we use this optical-pumping-spectroscopy method to determine isotope shifts on the narrow-line transition.

Next, we examine the transfer of optically pumped atoms from the spherical quadrupole magnetic trap to the narrow-line MOT formed by a single vertical beam.
To highlight this transfer, after extinguishing the broad-line MOT, we expose the atoms to a fixed-frequency narrow-line vertical beam and, over the next \qty{25}{\ms}, ramp down its intensity from $\qty{500}{\upmu \watt/\centi\meter^2}$ ($s_z\approx 220$) to $\qty{215}{\upmu \watt/\centi\meter^2}$ ($s_z\approx 90$) while also ramping down $B^\prime$ from \qty{24}{\gauss/\cm} to \qty{6}{\gauss/\cm}. 
This ramp has the effect of cooling down and isolating the atoms in the MOT while displacing them away from the magnetic trap so they can be separately imaged.

Images of the atom cloud after \qty{25}{\ms} of exposure to a single narrow-line MOT beam (Fig.~\ref{fig:optical_pumping}(b)) show that, for blue detuning, the atoms separate into two distributions.
One population of atoms remains localized at the center of the spherical quadrupole magnetic field.
These are atoms occupying weak-field-seeking $m_J$ states that are magnetically trapped.
The remaining population is now concentrated at a location physically below the magnetic trap.
These are atoms that have been captured in the single-beam narrow-line MOT.
We find the location of the MOT along $z$ to vary linearly with $\delta$, as shown in Fig.~\ref{fig:optical_pumping}(c). 

\begin{figure}[t]
    \centering
    \includegraphics{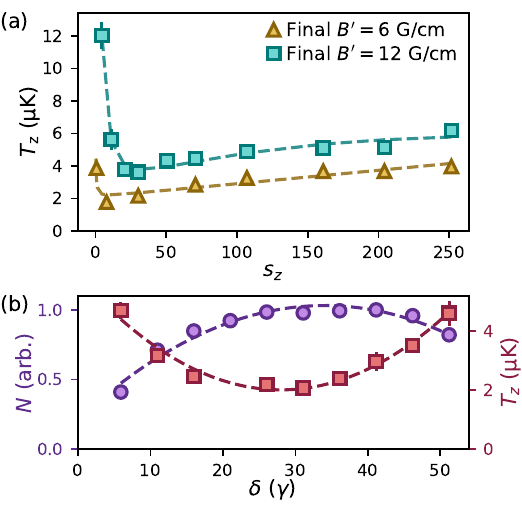}
    \caption{Characterization of the single-beam narrow-line MOT.
    (a) Temperature along the $z$ axis of atoms in the narrow-line MOT versus final saturation parameter of the vertical narrow-line cooling beam.
    (b) Atom number $N$, normalized to the maximum value, estimated by fitting a Gaussian profile to fluorescence images (purple circles), and temperature along the z-axis (red squares), versus detuning of the vertical $\lambda=\qty{1040}{\nm}$ beam.
    The dotted lines in both (a) and (b) are guides to the eye.
    }
    \label{fig:vertical}
\end{figure}

Finally, we examine the effectiveness of the single-beam narrow-line MOT in cooling the atomic gas.
After switching off the broad-line MOT, with the narrow-line beam at high intensity ($s_z = 220$) and the spherical quadrupole field at a high initial gradient ($B^\prime_z = 24$ G/cm), we ramp down the light intensity and field gradient over the next 25 ms.
The high initial values of $s_z$ and $B^\prime_z$ facilitate optical pumping and the transfer of atoms into the MOT, while the low final values facilitate cooling to low temperature.
The final atom number $N$ and one-dimensional (vertical) temperature $T_z$ are shown in Fig.\ \ref{fig:vertical}. 
Temperature is measured by time-of-flight (TOF) expansion.
The magnetic field gradient and MOT beams are extinguished and atoms freely expand for a variable duration before the atomic distribution is imaged by fluorescence on the $\lambda=\qty{498}{\nm}$ transition.
For a thermal gas with a Gaussian density distribution, the cloud size evolves according to
$\sigma(t)^2=\sigma_0^2+v_{\mathrm{rms}}^2 t^2$,
where $\sigma(t)$ is the Gaussian width of the atomic cloud after an expansion time $t$, $\sigma_0$ is the initial cloud size, and $v_{\mathrm{rms}}$ is the one-dimensional root-mean-square thermal velocity.
The temperature is then obtained from $v_{\mathrm{rms}}=\sqrt{k_B T/m}$.

The relationship between the temperature of the atoms in the narrow-line MOT and the optical intensity (parameterized by $s_z$) is qualitatively different from that of typical red-detuned six-beam MOTs.
To illustrate this, we consider a simple model of the temperature of the atoms confined to be along the $z$ axis.
For a given $B'$, atoms are trapped at equilibrium at a location where the scattering rate $R = F_B/\hbar k$ is constant.
Here, we neglect small contributions from gravity and variations of the magnetic force with saturation.
The heating rate from optical absorption and emission and magnetic force fluctuations, both being proportional to $R$, is constant, while the viscosity coefficient, $\alpha = - \partial_{v_z} F_z|_{v_z = 0}$, varies as
\begin{equation}
    \alpha\approx2\hbar k^2\frac{\sqrt{(F_\gamma/F_B-1)s_z-1}}{(F_\gamma/F_B)^2\, s_z}
\end{equation}
where $v_z$ is the vertical velocity. 
For $s_z\gg1$, power broadening of the narrow line cooling transition leads to $\alpha\propto1/\sqrt{s_z}$ so that $T_z$ decreases with decreasing $s_z$.
At lower $s_z$ as one approaches the condition  $F_\gamma\, s_z/(1+s_z) = F_B$,  the local detuning from resonance at the equilibrium MOT position approaches zero and the viscosity coefficient drops, causing the temperature to rise rapidly. 
Additionally, for higher $B'$, the higher scattering rate $R$ that balances the magnetic force generally increases the gas temperature at constant $s_z$.  The data in Fig.\ \ref{fig:vertical}(a) match this description qualitatively. 
A more quantitative model for the equilibrium $T_z$  requires consideration of the effects of off-axis motion, which are beyond the scope of this manuscript.
We find the lowest single-beam narrow-line MOT temperature along the vertical axis, $T_z=\qty{1.10(4)}{\upmu \kelvin}$, at a final gradient of $B'=\qty{3}{\gauss / \cm}$ and intensity in the vertical beam of $I_z=\qty{15}{\upmu \watt/\cm ^2}$ ($s_z\approx 7$). 

Fig.~\ref{fig:vertical}(b) illustrates the effects of detuning on the single-beam narrow-line MOT.
At detunings around $\delta \simeq 30 \, \gamma$ we observe a maximum transfer efficiency into the narrow-line MOT and a minimum in the vertical temperature.  
The increase in temperature at near detuning may be due to off-resonant $\pi$ and $\sigma^-$ transitions that introduce strong magnetic force fluctuations and heat the gas.  
It is less clear why we also observe higher temperatures at large ($>40\,\gamma$) detuning; this temperature increase may be due to the finite beam diameter of the vertical beam and the weaker radial magnetic forces at large distance from the magnetic trap center. 

\subsection{Optimization of the five-beam MOT}
\label{sec:5beam}

The single-beam MOT yields an atom cloud that is highly ansiotropic: strongly compressed and cooled along the vertical direction, but weakly compressed and at higher kinetic temperature in the radial directions.
For instance, at the settings described above that achieve the lowest $T_z$ temperatures, we simultaneously measure a kinetic temperature of $T_x = \qty{10(1)}{\upmu \kelvin}$ along the imaged horizontal axis.
This radial temperature is lower than the initial temperature of atoms in the broad-line MOT (around $\qty{100}{\upmu K}$), with some cooling achieved owing to coupling between axial and radial motion, but still much higher than $T_z$.

\begin{figure}[t]
    \centering
    \includegraphics[width=\columnwidth]{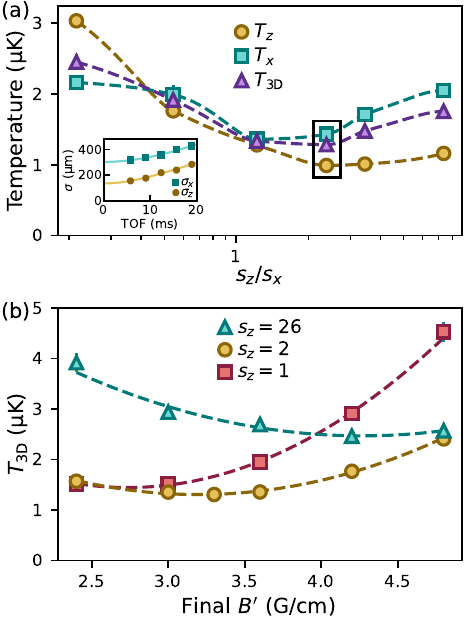}
    \caption{Temperatures in the five-beam narrow-line MOT following a 60\,ms 3D cooling ramp.
    (a) Vertical, transverse, and 3D temperatures versus ratio of vertical to transverse single-beam saturation parameter $s_z/s_x$ at fixed total saturation $s_z+4s_x\approx 7$ and final gradient 3.3\,G/cm.
    Note that only $1/4$ of the transverse beam intensity is of the correct polarization to drive the $\sigma^+$ transition for atoms on the $z$ axis.
    The inset shows the vertical and transverse Gaussian widths, $\sigma_z$ and $\sigma_x$, versus TOF expansion time for the lowest-temperature data at $s_z/s_x=2.4$. 
    Solid lines show the fit function used to derive temperature.
    (b) 3D temperature versus final magnetic gradient for several final saturation parameters with the ratio of vertical to horizontal single-beam saturation parameters ${s_z}/{s_x}$ fixed to 1.5 following the transverse cooling ramp.
    Higher saturation generally leads to reduced Doppler cooling and increased final temperatures.
    Dashed lines provide a guide to the eye.
    }
    \label{fig:temperature-plots}
\end{figure}

In order to cool and compress the atoms further radially, we next impose a five-beam narrow-line MOT.
After ramping the gradient and vertical intensity to intermediate values (${B'}=\qty{6}{\gauss / \cm}$ and $I_z=\qty{140}{\upmu \watt / \cm ^2}$), we turn on two pairs of counter-propagating $\qty{1040}{\nm}$ beams in the horizontal $x$ and $y$ directions.
A position-dependent scattering force emerges from the variation in the direction of the magnetic field: moving away from the $z$ axis, the orientation of the magnetic field tilts radially inward, increasing the inward scattering rate from left-hand circularly polarized MOT light on the near-resonant $m_J=5\to m_J=6$ transition.
This position-dependent force compresses the gas radially, reducing the transverse width by a factor of two compared to the single-beam MOT, while the locally red-detuned horizontal beams also provide radial Doppler cooling.

The inset of Fig.~\ref{fig:temperature-plots}(a) highlights the time-of-flight determination of temperatures at one setting, for which we find
$T_z=\qty{990(20)}{\nano \kelvin}$ and $T_x=\qty{1.43(10)}{\upmu \kelvin}$, respectively, giving a three-dimensional temperature of $T_{\mathrm{3D}}=(T_z+2T_x)/3=\qty{1.28(7)}{\upmu \kelvin}$.
Varying the power ratio between horizontal and vertical beams distributes kinetic energy variably between the axial and radial directions as shown in Fig.~\ref{fig:temperature-plots}(a), with increased power in the $z$ or $x$ axis laser beams leading to relatively lower temperatures along that direction.
The dependence of the five-beam MOT $T_{\mathrm{3D}}$ on the optical intensity (parameterized by $s_z$ at constant $s_z/s_x$) and $B'$ is similar to that of $T_z$ in the single-beam MOT, as shown in Fig.~\ref{fig:temperature-plots}(b).
At a given $B'$, both very low and very high $s_z$ reduce the viscosity coefficient along $z$ and lead to higher temperatures, while higher values of $B'$ induce higher optical scattering rates and thus higher temperatures.

\subsection{Performance of the narrow-line MOT}
\label{sec: performance}

Fig.~\ref{fig:E-levels-nMOT-cartoon}(d) depicts an experimental sequence optimized to maximize the transfer efficiency of atoms from the broad-line MOT to the final narrow-line MOT and to achieve the lowest final three-dimensional temperatures, assimilating the lessons learned in Sections \ref{sec:vertical} and \ref{sec:5beam}.
The detuning of the narrow-line MOT beams is constant at $\delta=23\,\gamma$ throughout the sequence.

About 25\% of the 165,000 atoms in the initial broad-line MOT~\cite{broadAtomNumberfootnote} are loaded into the narrow-line MOT in the early stages of the experimental sequence, and no atom loss is observed throughout the single-beam and five-beam intensity and gradient ramps, leading to a final narrow-line MOT atom number of about 40,000~\cite{narrowAtomNumberFootnote}.

The narrow-line MOT transition in Ti is not fully closed.
Atoms in the excited $\mathrm{z ^5G_6^o}$ state can also decay on spin-forbidden but dipole-allowed  single-photon transitions to the $\mathrm{a ^3G_5}$ state.
The branching ratio for decays from the excited state to states other than the laser cooling state is estimated theoretically to be $\alpha = 3.1(2)\times 10^{-7}$ \cite{eustice_optical_2023}.  
The observed lifetime of the narrow-line MOT, $\tau_\mathrm{nMOT} = 472(11)$ ms,  allows us to establish a rough upper bound on this branching ratio $\alpha$.
Assuming atom loss from the narrow-line MOT is caused entirely by leakage decay of the excited state, with the assumption that atoms in states outside the laser-cooling transition are lost, and estimating the scattering rate at the final conditions of the narrow-line MOT sequence as $R\approx\qty{5e3}{\per\second}$ , we obtain $\alpha < 1/(R \, \tau_\mathrm{nMOT}) = 4.4(1) \times 10^{-4}$.
The lifetime of the narrow-line MOT is in fact similar to that of magnetically trapped atoms in the absence of resonant light ($\tau_\mathrm{mag} = 464(52)$\,ms), suggesting that the loss of atoms from the narrow-line MOT is mostly caused by background gas collisions in the experimental vacuum chamber and is compatible with a much smaller value of $\alpha$ than indicated by the experimental upper bound.

We measure Gaussian rms widths of $\sigma_z=\qty{75}{\upmu \meter}$ in the vertical direction of the narrow-line MOT and $\sigma_x=\qty{240}{\upmu \meter}$ in the horizontal directions, assuming cylindrical symmetry about the $z$ axis.
These measured widths may be broadened by diffusion and the presence of residual magnetic field gradients during the $\qty{50}{\upmu \s}$ imaging pulse, but allow us to place an upper bound on the narrow-line MOT size~\cite{densityFootnote}.
The lower bound on final density of the narrow-line MOT is then $5.9(5)\times10^8$\,cm$^{-3}$, leading to a phase space density of $1.9(2)\times10^{-5}$ at the minimum temperature $T_{\mathrm{3D}}=\qty{1.28(7)}{\upmu \kelvin}$. 
For previously observed narrow-line MOTs of dipolar atoms, heating due to dipolar relaxation was cited as a possible mechanism that could limit the achievable temperature and phase space density \cite{lu_strongly_2011}.
However, dipolar relaxation is not expected to limit the temperature at our observed densities; see Appendix \ref{app:dipolar} for calculations of dipolar heating rates. 

\section{Spectroscopy of the narrow line}
\label{sec:spectroscopy}

Having realized a narrow-line MOT of titanium, we are able to characterize the $\mathrm{3d^3(^4F)4s\,a^5F_5} \rightarrow \mathrm{3d^2(^3F)4s4p(^3P^o) \, z^5G_6^o}$ transition itself.
We measure the spin polarization of the gas, the isotope shifts on the bosonic isotopes $^{50}$Ti and $^{48}$Ti, and the lifetime of the excited $\mathrm{z^5G_6^o}$ state. 

\subsection{Stern-Gerlach analysis of ultracold Ti gas}\label{sec:SG}

The $m_J$ distribution of ultracold metastable-state Ti is readily probed using Stern-Gerlach separation.
We perform Stern-Gerlach analysis on atoms cooled to $\sim \qty{1}{\upmu \kelvin}$ to determine the spin polarization of narrow-line MOT atoms and to measure the offset of the narrow line zero-field atomic resonance from the resonance of the ULE cavity to which the 1040-nm-wavelength laser is frequency locked. 

Atoms in the blue-detuned narrow-line MOT are expected to be polarized in the stretched $m_J=+5$ state due to energetic optical pumping as described in Sec.~\ref{sec:theory}.
To find the $m_J$ population distribution, a narrow-line MOT is loaded using the optimized sequence described in Sec.~\ref{sec:optimization}.
Narrow-line beams are extinguished and the magnetic field gradient is rapidly (with $\sim\qty{250}{\upmu \second}$ response time) increased to \qty{24}{G/cm} for \qty{2}{\milli\second}. 
The nonzero $B_z$ and high gradient $B'$ of the spherical quadrupole field at the location of the detuned narrow-line MOT provides a strong vertical magnetic field gradient that imparts an $m_J$-dependent impulse to the atoms.
The gradient field is then turned off and the atoms propagate freely for a \qty{10}{\milli\second} time of flight, causing the different $m_J$ atom populations to separate from one another.
The atom distribution is then imaged by fluorescence in a $\qty{500}{\upmu \second}$ pulse of resonant 498-nm-wavelength light.
The fractional atom population in the $m_J=+5$ sublevel is measured as $>98\%$, and is limited by our ability to resolve distinct $m_J$ distributions. 

\begin{figure}[t]
    \centering
    \includegraphics{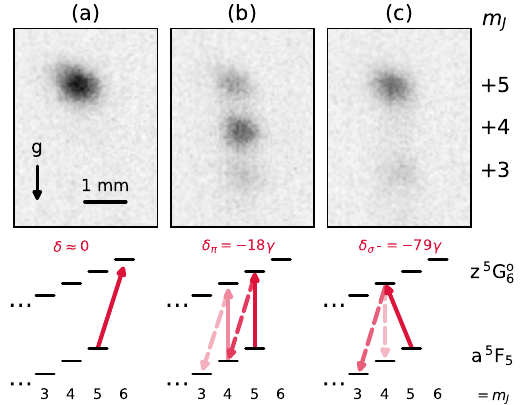}
    \caption{Stern-Gerlach separation of ultracold Ti atoms.
    As the depumping pulse frequency is scanned, we observe (a) atoms polarized in the $m_J=+5$ state when the pulse is off-resonant, (b) occupation of the $m_J=+4$ and $m_J=+3$ states when addressing the $\pi$ transition, and (c) occupation of $m_J=+3$ when addressing the $\sigma^-$ transition.
    Schematics of the driven transitions, noting their detuning from the zero-field resonant frequency, are shown below the Stern-Gerlach image.
    Excitations are shown in sold lines and decay channels in dashed lines.
    For decay and subsequent excitation pathways, the darkness of the line is scaled to the the transition probability according to Clebsch-Gordon coefficients.
    }
    \label{fig:stern-gerlach}
\end{figure}

A spin mixture can be created prior to Stern-Gerlach separation by applying a short depumping pulse to the atoms using the transverse narrow-line beams before increasing the magnetic gradient and releasing the atoms.
By tuning the transverse beam frequency during this brief pulse to either the $\pi$ or the $\sigma^-$ optical transition frequencies, matching the Zeeman shifts at the non-zero magnetic field present at the location of the narrow-line MOT, we observe depopulation of the $m_J=+5$ state and population of lower magnetic sublevels; see Fig.\ \ref{fig:stern-gerlach}.

This optical depumping signal allows us to locate the zero-magnetic-field narrow-line resonance frequency with higher accuracy than the optical pumping method described in Sec.\ \ref{sec:vertical}.
As described further in Appendix \ref{app:SG}, we control the magnetic field at the location of the narrow-line MOT by varying the detuning of the narrow-line-MOT beams, and then use Stern-Gerlach analysis to measure the depumping transition frequencies at that magnetic field.
The measured frequencies are extrapolated to zero magnetic field to determine the zero-field transition frequency.
All frequency detunings reported in this work are referenced to this Zeeman-spectroscopy determination of the zero-field resonance.

\subsection{Isotope shifts}
\label{sec:isotope}

\begin{figure}[t]
    \centering
    \includegraphics[width=\columnwidth]{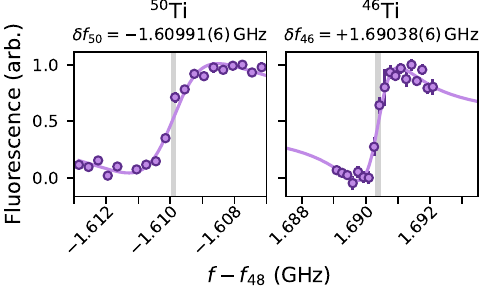}
    \caption{Isotope shift spectroscopy on the bosonic isotopes $^{50}$Ti and $^{46}$Ti.
    Fluorescence on the $\lambda=\qty{498}{\nm}$ line is shown as the frequency offset of the $\qty{1040}{\nm}$ light from the $^{48}$Ti resonance $f_{48}$ is varied. A dispersive lineshape of the form given in Eq.~\ref{eq:fit} fits the data (purple line), with light gray bars indicating the line center and the $1\sigma$ uncertainty in the fit values. 
    The isotope shifts determined from these fits are reported above.
    }
    \label{fig:isotopes}
\end{figure}

We measure the isotope shifts of the two other bosonic isotopes of titanium, $\mathrm{^{46}Ti}$ and $\mathrm{^{50}Ti}$, using optical pumping spectroscopy: Locking the laser near the expected isotope shift, we determine the transition frequency using the quadrupole magnetic trap that is present for Ti atoms after turning off the \qty{498}{\nm} broad-line MOT light. 
Atoms are held in the magnetic trap for \qty{50}{\ms}, with the gradient of the magnetic trap ramped from 24\,G/cm to 12\,G/cm over the first 25\,ms, subject to a single vertical beam of $\qty{1040}{\nm}$ light at an intensity of $\qty{500}{\upmu \watt/\centi\meter^2}$ ($s_z\approx 220$).
We then count the remaining atom number by fluorescence imaging.  By scanning the narrow-line light frequency, we obtain spectra (Fig.~\ref{fig:isotopes}) similar to those seen for $^{48}$Ti (Fig.~\ref{fig:optical_pumping}).
Fitting the line centers of these spectra quantifies the isotope shifts on the narrow line transition, relative to the $^{48}$Ti resonance,  to be $\delta f_{50}=-1.60991(6)$\,GHz and $\delta f_{46}=+1.69038(6)$\,GHz for the $^{50}$Ti and $^{46}$Ti isotopes, respectively.
At blue detuning, we observe the formation of single-beam narrow-line MOTs for both $^{46}$Ti and $^{50}$Ti.
These narrow-line MOTs have similar characteristics as those of $^{48}$Ti, although we restricted more quantitative analysis, as discussed above, to $^{48}$Ti owing to its larger natural abundance and MOT atom numbers.

\subsection{Lifetime of the $\mathrm{z^5G_6^o}$ state}
\label{sec:lifetime}

We measure the lifetime of the $\mathrm{z^5G_6^o}$ state for $^{48}$Ti via state-shelving methods in both the broad-line and narrow-line MOTs.
By applying a $\qty{200}{\upmu \s}$, $\qty{8}{\milli \watt / \cm ^2}$ pulse of resonant 1040-nm-wavelength  light onto atoms that are actively trapped in the broad-line MOT, a large fraction of atoms near the center of the MOT ($\sim 20$\% of the total MOT population) are excited to the $\mathrm{z^5G_6^o}$ state and no longer scatter broad-line MOT light until they decay back to the $\mathrm{a^5F_5}$ state.
By collecting the 498-nm-wavelength fluorescence onto a photomultiplier tube (PMT), we measure the decay of the atoms shelved in the $\mathrm{z^5G_6^o}$ state as an increase in the \qty{498}{\nm} photon count rate.
We fit an exponential decay model to the data and find an excited state lifetime of $\tau_b=\qty{9(2)}{\upmu \s}$. 

\begin{figure}[t]
    \centering
    \includegraphics{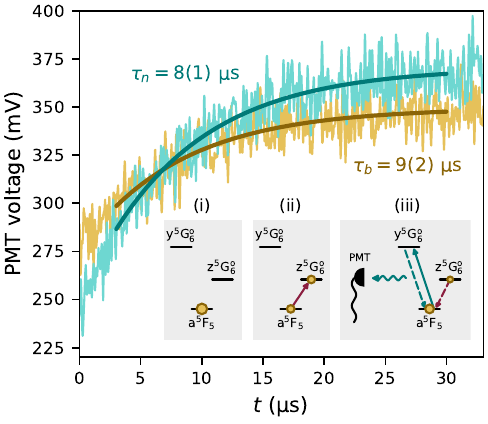}
    \caption{The average photon scattering rate, $R$, on the $\lambda=\qty{498}{\nm}$ line, measured by a photomultiplier tube (PMT), is decreased when atoms are subject to $\lambda=\qty{1040}{\nm}$ light due to shelving into the $\mathrm{z^5G_6^o}$ state, and recovers when the \qty{1040}{\nm} light is turned off.
    The inset illustrates the population dynamics during shelving spectroscopy: Initially, atoms are in the $\mathrm{a^5F_5}$ state (i).
    Next, $\qty{1040}{\nm}$ light shelves atoms in the excited $\mathrm{z^5G_6^o}$ state (ii).
    The rf input to the \qty{498}{\nm} probe light AOM is then switched on ($t=\qty{-5}{\upmu \second}$) and the rf input to the \qty{1040}{\nm} shelving light AOM switched off ($t=0$).
    Atoms decay from $\mathrm{z^5G_6^o}$ to $\mathrm{a^5F_5}$, and can then scatter $\lambda=\qty{498}{\nm}$ light which is collected by a PMT (iii). 
    The PMT voltage trace is shown both for atoms in a broad-line MOT following an intense $\lambda=\qty{1040}{\nm}$ pulse (gold) and for atoms initially trapped in a narrow-line MOT (teal). The solid lines show the fit of the data to an exponential decay curve, with $\tau_b=\qty{9(2)}{\upmu \s}$ using the broad-line MOT method and $\tau_n=\qty{8(1)}{\upmu \s}$ using the narrow-line MOT method. 1280 (1408) measurements were averaged for the determination of $\tau_b$ ($\tau_n$).
    }
    \label{fig:spectroscopy}
\end{figure}

We apply a similar method starting with atoms loaded into a narrow-line MOT with total saturation parameter $s\approx50$.
After the loading and cooling ramp, $\sim33$\% of the atoms are shelved in the excited state. We switch on the \qty{498}{\nm} light and, after a $\qty{5}{\upmu \second}$ delay to account for acousto-optic switching times, turn off the \qty{1040}{\nm} light.
We observe an increase in \qty{498}{\nm} scattering, and by fitting this signal, we determine a lifetime of $\tau_n=\qty{8(1)}{\upmu s}$.

Averaged oscilloscope traces of the signal from both methods are shown in Fig.~\ref{fig:spectroscopy}, and a schematic of the shelving method (relevant to measurements in both the broad-line and narrow-line MOTs) is shown as an inset.
In both cases, the atomic distribution is not expected to move significantly over the timescale of $\qty{<10}{\upmu \s}$. 
The two methods expose the $\mathrm{z^5G_6^o}$ atoms to significantly different \qty{498}{\nm} optical powers and atom densities, and yet they agree closely in the measured lifetime.
For this reason, we believe the observed $\qty{8}{\upmu \s}$ lifetime is not affected by light-induced decay of $\mathrm{z^5G_6^o}$ atoms from the \qty{498}{\nm} wavelength light; neither one-body processes such as off-resonant atomic excitation nor two-body processes such as light-induced collisional decay.
Combining the results of these two measurements in a uncertainty-weighted average, we determine the lifetime of the $\mathrm{z^5G_6^o}$ state to be $\tau=\qty{8.2(9)}{\us}$, with the corresponding transition linewidth of of $\gamma/2\pi=20(2)$\,kHz.
Our linewidth measurement is consistent with previous semi-empirical and theoretical predictions of 21.9\,kHz \cite{KuruczAtomicDatabase} and 16.4(1.3)\,kHz \cite{eustice_optical_2023}, respectively, and disagrees with a result of 11.3(1.6)\,kHz derived from the solar intensity spectrum \cite{borrero_accurate2003}. 
Throughout this text, calculations of parameters such as the saturation intensity and normalized detuning ($\delta/\gamma$) are based on our experimentally measured linewidth.

\section{Conclusion}
\label{sec:conclusion}

We have demonstrated a narrow-line magneto-optical trap of titanium atoms using the $\mathrm{a^{5}F_5}\rightarrow\mathrm{z^{5}G_6}$ transition at \qty{1040}{\nm}, achieving $\sim \qty{1}{\upmu \kelvin}$ temperatures while simultaneously polarizing atoms in the stretched $m_J=+5$ state.
The large magnetic moment of the Ti metastable state enables the narrow-line trap to operate in a regime where magnetic forces play an essential role in the trapping dynamics, providing a distinct platform compared with narrow-line MOTs in non-magnetic atoms.
Our detailed characterization of the properties of the narrow-line MOT is restricted to the most abundant isotope, $^{48}$Ti, due to the higher atom numbers available; however, single-beam narrow-line MOTs of the other bosonic isotopes, $^{46}$Ti and $^{50}$Ti, were loaded using the same intensity and gradient sequence described in this work. 

The narrow-line MOT provides a foundation for producing dense ultracold Ti samples. In the current setup, direct loading of the broad-line MOT from the output of a Ti sublimation source yields Ti MOTs with modest atom number; here, we perform experiments with atom numbers around $10^5$, while previous experiments operated with somewhat higher numbers ($\sim10^6)$~\cite{eustice_magneto-optical_2025}.
The integration of conventional methods to improve the loading rate into the broad-line MOT, such as Zeeman slowing, is expected to increase the broad-line MOT atom number by orders of magnitude, which should translate into narrow-line MOTs with correspondingly larger atom numbers.
In addition, the observed 25\% loading fraction into the narrow-line MOT is limited by optical pumping efficiency and may be improved by optically pumping with a polarization mixture to address atoms at any position within the magnetic trap, or introducing a short uniform-field optical pumping stage. 
At the higher narrow-line MOT densities that may result, heating due to inelastic dipolar collisions might limit the narrow-line MOT temperature, as observed for Dy narrow-line MOTs \cite{lu_strongly_2011}.
This limitation can be mitigated through rapid adiabatic passage to the $m_J=-5$ state and transfer to a red-detuned narrow-line MOT operating on the $\sigma^-$ transition. 

With the higher atom numbers expected from experimental upgrades, narrow-line cooling offers a shortcut to quantum degeneracy. 
Sr has been laser cooled directly to degeneracy in $\qty{100}{\ms}$ using a narrow line~\cite{stellmer_laser2013}, while narrow-line MOTs similar to the one described in this work facilitated Bose condensation of Er on a sub-second timescale \cite{phelps_subsecond2020} as well as Dy \cite{lu_strongly_2011}.
 The narrow-line transition also provides a pathway for direct laser cooling of Ti atoms to low temperatures in optical traps, including tweezer traps.
For complex atoms such as Ti and the lanthanide elements, the illumination of atoms in optical traps typically leads to strongly state-dependent potentials owing to large vector and tensor ac Stark shifts.
Such anisotropic optical polarizabilities generally interfere with polarization-gradient cooling, which is commonly employed in optical tweezer trapping of simpler elements such as alkalis.
However, as has been shown in Dy \cite{chalopin_anisotropic2018}, one can realize a magic-wavelength optical trap on a stretched-state narrow-line optical transition.
For the case of Ti, one would employ trapping light of a specific wavelength and polarization such that the  $\mathrm{a ^5F_5} \rightarrow \mathrm{z ^5G_6^o}$ transition between the $m_J=\pm5$ and $m_{J^\prime}=\pm6$ sublevels is unshifted by optical trapping light.
For optical tweezer traps or optical lattice potentials with trapping frequencies on the order of 100 kHz, Doppler-limited narrow-line cooling would be sufficient to cool Ti atoms to the motional ground state.
Such ground-state cooling could facilitate high-fidelity fluorescence imaging in a quantum gas microscope~\cite{Cheuk_quantum2015,Parsons_site2015}, or prepare atoms deep in the Lamb-Dicke regime with respect to the telecommunications-band clock transition in Ti at \qty{1549}{nm}, enabling high-performance clock operation~\cite{eustice_optical_2023}.

\acknowledgements
This material is based upon work supported by the U.S. Department of Energy, Office of Science, National Quantum Information Science Research Centers, Quantum Systems Accelerator.  Additional support is acknowledged from the the ARO (Grants No.\ W911NF-23-1-0244, and No.\ W911NF-26-1-A209), the NSF QLCI program through Grant No.~OMA-2016245).  R.\ D.\ is supported through the Air Force Office of Scientific Research under award number FA9550-23-F-0014 in the amount of \$134,600.

\appendix

\section{Stern-Gerlach spectroscopy of the narrow line}
\label{app:SG}

Stern-Gerlach spectroscopy is used to determine the offset of the narrow line zero-field atomic resonance from the optical resonance frequency of our ULE optical cavity to which the 1040-nm-wavelength emission of our laser light source is stabilized.
For this spectroscopy, a narrow-line MOT is loaded using the optimized sequence described in Sec.\ \ref{sec:optimization}, at a detuning $\delta$ from the narrow-line resonance frequency that we vary to control the position of the narrow-line MOT within the applied spherical quadrupole field.
The narrow-line MOT beams are then turned off, and a short ($\qtyrange{5}{30}{\upmu \second}$), 3\,mW ($4s_x \sim 1300$) depumping pulse is applied to the atoms using the transverse 1040\,nm beams.
The magnetic field gradient is then increased to \qty{24}{G/cm} for \qty{2}{\milli\second} and turned off to release the atoms for a \qty{10}{\milli\second} time of flight and fluorescence imaged with a $\qty{500}{\upmu \second}$ pulse of \qty{498}{\nm} light.

The narrow-line MOT forms at different magnetic fields (i.e.\ positions along the $z$ axis) for different detunings of the MOT beam.
At each field, we scan the depumping pulse frequency across the $\pi$ and $\sigma^-$ transitions.
At either transition, we observe depopulation of the $m_J = +5$ state and a population of atoms in states with lower values of $m_J$.
The depumping dynamics on the $\pi$ and $\sigma^-$ transitions are somewhat different, as indicated by the schematic level diagrams in Fig.\ \ref{fig:stern-gerlach}.
For the $\pi$ transition, excitation of the initial $m_J = +5$ population to the excited $m_{J^\prime} = +5$ sublevel, which occurs when the detuning of the depump beam matches $\delta_\pi = (m_J g_e - m_J g_g) \mu_B B = - (m_J/15) \mu_B B$, depumps atoms into the $m_J = +4$ state.
The weak $m_J$ dependence of the transition Zeeman shift typically allows the $\pi$ transitions for atoms in the $m_J = +4$ sublevel also to be near resonance, permitting a sequential depumping of atoms to the $m_J = +3$ level.
At the $\sigma^-$ depumping resonance, we see an initial transfer of atoms from the $m_J = +5$ state, through the excited $m_{J^\prime} = +4$ state, to both the $m_J = +3$ and $m_J = +4$ sublevels, with the population in the $m_J = +3$ state being larger owing to the different values of Clebsch-Gordon coefficients that govern the decay from the excited state.
Sequential depumping occurs also on the $\sigma^-$ transitions, whose detuning from the zero-field resonance, $\delta_{\sigma^-} = ((m_J - 1) g_e - m_J g_g) \mu_B B = -(4/3 + m_J/15) \mu_B B$, also varies only weakly with $m_J$.
In both cases, we determine the resonant depumping frequency as that which depletes the $m_J = +5$ population most strongly.

The magnetic field at the location of the atoms can be inferred from the narrow-line MOT beam detuning.
However, the slight offset between the position of the Zeeman-tuned resonance $z_\delta$ and the equilibrium position of the atoms makes this inference somewhat imprecise.
Thus, we identify the zero-field resonance condition by plotting the two depumping resonance frequencies $\delta_\pi$ and $\delta_{\sigma^-}$, against one another (Fig.~\ref{app:SG}), and identifying thereby the optical resonance frequency (as measured against the frequency of our ULE cavity) at which the $\pi$ and $\sigma^-$ transitions occur at the same depumping beam frequency; in other words, where the Zeeman shift goes to zero.
With this zero-field frequency, we can then translate between optical frequency offsets from our ULE stabilization cavity to the detunings $\delta$ from the narrow-line resonance that are reported throughout this work.

\begin{figure}[t]
    \centering
    \includegraphics{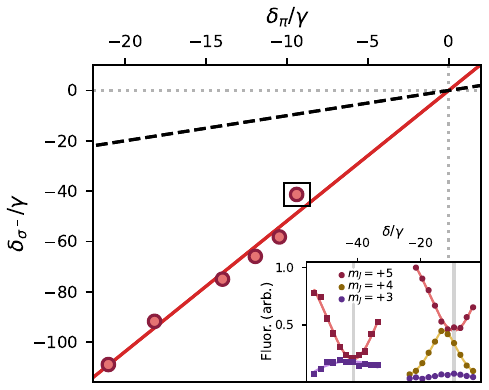}
    \caption{Stern-Gerlach spectroscopy on the narrow line.
    The $\sigma^-$ resonant frequency (red) is plotted as a function of the $\pi$ resonant frequency measured at the same magnetic field. 
    The zero-field resonance of the narrow-line transition is extrapolated by finding the frequency at which the line fit to the $\delta_{\sigma^-}$ data intersects the line $\delta_{\sigma^-}=\delta_\pi$.
    The inset shows normalized fluorescence at the positions of $m_J=+5$ (red), $m_J=+4$ (gold), and $m_J=+3$ (purple) separated clouds versus detuning of the depumping pulse from zero-field resonance at $B\approx 0.16$\,G.
    The resonant frequency (vertical gray lines) is determined as that which minimizes the $m_J=+5$ population. 
    }
    \label{fig:sg-appendix}
\end{figure}

\section{Calculation of dipolar relaxation effects}
\label{app:dipolar}

In previous reports on blue-detuned narrow-line MOTs of magnetic atoms, the observed temperature was above the Doppler limit that was expected for those transitions \cite{berglund_narrow_2008,lu_strongly_2011}.
One of the proposed mechanisms for this discrepancy was additional heating due to dipole-dipole interactions of the atoms in the trap, with dipolar relaxation being the largest contribution to this heating mechanism~\cite{lu_strongly_2011}.
To consider whether this is likely to be a significant effect in our system, we estimate the dipolar relaxation heating rate and compare it to the expected heating rate from photon scattering.

\begin{figure}
    \centering\includegraphics{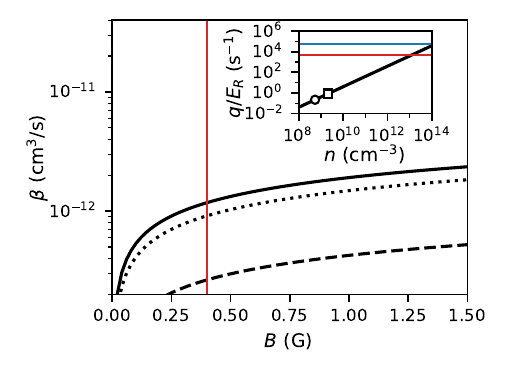}
    \caption{Dotted, dashed, and solid lines show the dipolar relaxation rate constants $\beta_1$, $\beta_2$, and $\beta=\beta_1+\beta_2$ for a gas of Ti atoms in the $\mathrm{a^5F_5}$ $m_J=5$ sublevel at a temperature of $\qty{1}{\upmu \kelvin}$ as a function of $B$.
    The vertical red line indicates the typical $B\approx\qty{0.4}{G}$ experienced by atoms at the end of our five-beam narrow-line MOT sequence.
    The inset shows the heating rate from dipolar collisions $q$ for atoms in a \qty{0.4}{G} field as a function of density $n$.
    The maximum density observed in the experiment via fluorescence imaging is indicated with a circle, while the maximum density in our system estimated from a calculation~\cite{densityFootnote} is indicated with a square.
    The horizontal red and blue lines indicate the heating rate expected from optical scattering of narrow-MOT light at the scattering rate of atoms in the MOT (\qty{5e3}{\per\s}) and at $\gamma/2=\qty{63e3}{\per\s}$.
    }
    \label{fig:dipolar_relaxation_calc}
\end{figure}

The two-body dipolar relaxation constants for single- and double spin flip ($\beta_{1,2}$) is calculated for a spin polarized $m_J=J$ gas as a function of the magnetic field at the collision location following~\cite{chom23review} and is plotted in Fig.~\ref{fig:dipolar_relaxation_calc}.
These coefficients are in turn related to the inelastic collision rate for each process by the local atomic density $n$ as $R_{1,2}=n\beta_{1,2}$.
Each spin flip collision transfers the energy stored in the Zeeman shift of the atomic spins into kinetic energy, resulting in a total heating rate of $q=\sum_{i=1,2}ig_JB\mu_\mathrm{B}R_i$ (shown for experimentally relevant fields in the inset to Fig.~\ref{fig:dipolar_relaxation_calc}).
If the rate of collisions is sufficiently small that any atoms that scatter to $m_J<J$ states are rapidly pumped back to the stretched state, then this conversion from Zeeman energy to kinetic energy can effectively capture the effect of dipolar relaxation.
We find that for the temperatures and densities achieved in our narrow-line MOT, the heating rate from dipolar relaxation is several orders of magnitude lower than the heating rate from the scattering of narrow-line MOT light and is thus not significant in the system reported here.

\bibliography{references,footnotes}

\end{document}